%% file: main.tex
\documentclass[conference]{IEEEtran}
\IEEEoverridecommandlockouts
\makeatletter
\newcommand*\titleheader[1]{\gdef\@titleheader{#1}}
\AtBeginDocument{%
  \let\st@red@title\@title
  \def\@title{%
    \bgroup\normalfont\large\centering\@titleheader\par\egroup
    \vskip1.5em\st@red@title}
}
\makeatother

\usepackage{cite}
\usepackage{graphicx}
\usepackage{subfigure}
\usepackage{mathtools}
\usepackage{svg}
\usepackage{amssymb}

\usepackage{lipsum}

\usepackage{graphicx}
\newcommand{\R}{\mathbb{R}}
\newcommand{\N}{\mathbb{N}}

\def\BibTeX{{\rm B\kern-.05em{\sc i\kern-.025em b}\kern-.08em
    T\kern-.1667em\lower.7ex\hbox{E}\kern-.125emX}}

\title{An Investigation on Inherent Robustness of Posit Data Representation}
\titleheader{TO APPEAR IN VLSID 2021\vspace{-10mm}}
\author{\IEEEauthorblockN{Ihsen Alouani\IEEEauthorrefmark{1}, Anouar BEN KHALIFA\IEEEauthorrefmark{2}, Farhad Merchant\IEEEauthorrefmark{3}, Rainer Leupers\IEEEauthorrefmark{3}}
		\IEEEauthorblockA{\IEEEauthorrefmark{1}IEMN Lab CNRS UMR 8520, INSA Hauts-De-France}
        \IEEEauthorblockA{\IEEEauthorrefmark{2}National Engineering School of Sousse, University of Sousse, Tunisia}
        \IEEEauthorblockA{\IEEEauthorrefmark{3}Institute for Communication Technologies and Embedded Systems, RWTH Aachen University, Germany}
        Ihsen.Alouani@uphf.fr, anouar.benkhalifa@eniso.rnu.tn \{farhad.merchant, leupers\}@ice.rwth-aachen.de \vspace{-6mm}}

\begin{document}
\bstctlcite{IEEEexample:BSTcontrol} 
\maketitle

\begin{abstract}
\input{abstract}
\end{abstract}


\section{Introduction}
\input{introduction}

\section{Background and Related Works}\label{sec:background}
\input{background}

\section{Related Work}\label{sec:rw}
\input{rw}

\section{Proposed Methodology}\label{sec:prop}
\input{methodology}

\section{Experimental Setup and Results}\label{sec:exp}
\input{results}

\section{Conclusion}\label{sec:conc}

\input{conclusion}

\bibliographystyle{IEEEtran}
\bibliography{IEEEabrv,ref}

\end{document}

%% file: abstract.tex
As the dimensions and operating voltages of computer electronics shrink to cope with consumers’ demand for higher performance and lower power consumption, circuit sensitivity to soft errors increases dramatically. Recently, a new data-type is proposed in the literature called \emph{posit} data type. Posit arithmetic has absolute advantages such as higher numerical accuracy, speed, and simpler hardware design than IEEE 754-2008 technical standard-compliant arithmetic. In this paper, we propose a comparative robustness study between 32-bit posit and 32-bit IEEE 754-2008 compliant representations. At first, we propose a theoretical analysis for IEEE 754 compliant numbers and posit numbers for single bit flip and double bit flips. Then, we conduct exhaustive fault injection experiments that show a considerable inherent resilience in posit format compared to classical IEEE 754 compliant representation. To show a relevant use-case of fault-tolerant applications, we perform experiments on a set of machine-learning applications.
In more than $95\%$ of the exhaustive fault injection exploration, posit representation is less impacted by faults than the IEEE 754 compliant floating-point representation. Moreover, in $100\%$ of the tested machine-learning applications, the accuracy of posit-implemented systems is higher than the classical floating-point-based ones.

%% file: introduction.tex
As sub-micron technology dimensions sharply decrease to a few nanometer ranges in commercialized integrated circuits, the sensitivity of electronic circuits increases drastically \cite{nuclei}. Hence, embedded microprocessors are becoming more vulnerable to soft errors, and designing dependable systems is a challenging task for chip designers. In fact, these systems have to operate reliably even in the presence of faults, to sustain the present growth rate of device count and clock frequency with continuously growing reliability issues. Moreover, the sensitivity of chips is also intensified by voltage scaling~\cite{scaling}. Undesirable and accidental faults become more frequent in new generation computing systems where the systems are running heavy-duty numerical computations. A single-event upset (SEU) or multi-bit upset (MBU) could bring a catastrophic outcome in mission-critical applications such as space and missile-navigation applications. Hence, having reliable arithmetic that is resilient to errors is a primary requirement for mission-critical computing systems. 

\emph{Posit} is a new data type that is capable of storing more information-per-bit compared to its IEEE 754 compliant counterparts~\cite{john1}. For example, a 32-bit posit number can have a similar dynamic range and better accuracy at the same time compared to the 32-bit IEEE 754 compliant number. In general, m-bit posit has a higher dynamic range and better numerical accuracy properties compared to n-bit IEEE 754 compliant number where $m=n$. It is shown in the literature that for computing systems, $n$-bit IEEE 754 compliant numbers can be replaced by $m$-bit posit numbers where $m<n$ since the posit number system exhibits a trade-off between accuracy and dynamic range~\cite{Farhad9}~\cite{expand1}. These trade-offs allow the selection of the desired posit format that is suitable for computing systems without compromising accuracy and performance~\cite{clarinet1}~\cite{saxena1}. Further details of the number system and the formats are discussed in Section \ref{sec:posit_sub}. Reliability aspects of posit arithmetic are yet to be explored by the research community. 

To the best of our knowledge, this is the first comparative study on the inherent fault tolerance of posit arithmetic vis-\`a-vis its IEEE counterpart. We carry out an extensive investigation of reliability through an exhaustive fault injection scheme. The major contributions of the paper are as follows:
\begin{itemize}
    \item We propose a theoretical analysis and an exhaustive reliability exploration of  posit arithmetic vis-\`a-vis its IEEE-754 compliant counterparts.
    \item We conduct exhaustive reliability exploration as well as machine-learning (ML) benchmarks under fault injection.
    \item We show promising results in posit arithmetic that may encourage its utilization in safety-critical applications, as well as approximate computing. 
\end{itemize}

For the reproducibility, we make our framework open source~\cite{git_fault}. The rest of the paper is organized as follows: In Section \ref{sec:background} we present a background on posit arithmetic and soft errors followed by related work in Section \ref{sec:rw}. Section \ref{sec:prop} describes the analysis and the proposed methodology for error resilience using posit arithmetic. In section \ref{sec:exp}, experimental setup and results are discussed. We summarize our work in Section \ref{sec:conc}. 

%% file: background.tex
\subsection{IEEE 754 Compliant and Posit Number Systems}\label{sec:posit_sub}
\begin{figure}[!t]
\centering
\includegraphics[width=\columnwidth]{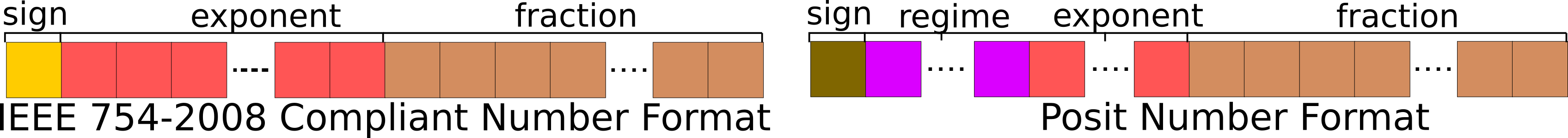}
\caption{Description of the IEEE 754 single-precision floating-point and \emph{posit} formats}
\label{fig:ieee754}
\end{figure}

The IEEE 754-2008 compliant floating-point format binary numbers are composed of three parts: a sign, an exponent and a fraction part (see Fig. \ref{fig:ieee754}). The sign is the most significant bit indicating whether the number is positive or negative. In a single-precision format, the following 8 bits represent the exponent of the binary number ranging from $-126$ to 127. The remaining 23 bits represent the fractional part. 
The normalized format of floating-point numbers is:
\begin{equation}\label{eqn:ieee754}
val = ( -1 )^{\textit{sign}} \times 2^{exp-bias} \times ( 1.fraction ) 
\end{equation}

\textit{Posit arithmetic} is proposed as a drop-in replacement for IEEE 754 compliant arithmetic in 2017 \cite{john1}. The posit number format has several absolute advantages over IEEE 754 compliant arithmetic such as higher accuracy, higher dynamic range, simpler hardware implementation for arithmetic operations, lower area and energy footprints~\cite{Gustafson2020}. Besides, it is shown in the literature that m-bit posit adders/multipliers can safely replace $n$-bit IEEE 754 compliant adders/multipliers where $m<n$ \cite{Farhad9}. Hence, posit representation confirms more information-per-bit compared to its IEEE 754 counterpart representation. Furthermore, with posit representation, there are no redundant representations and the overflow/underflow in the computations is nonexistent with posit arithmetic. The subnormal numbers are handled in a normal way with posit representation unlike IEEE 754 representation and there are only two exception cases: zero and not-a-real (NaR). For all other cases, the value $val$ of a posit is given by 
\begin{align}\label{eqn:posit}
	val  = & (-1)^{\textit{sign}} \times \textit{useed}^{k} \times 2^\textit{exp}\times (1+\sum_{i=1}^{\textit{fn}-1}b_{\textit{fn} - 1 - i}2^{-i})
\end{align}
The regime indicates a scale factor of $\textit{useed}^{k}$ where $\textit{useed} = 2^{2^\textit{es}}$ and \emph{es} is the exponent size. The numerical value of $k$ is determined by the \emph{run length} of 0 or 1 bits in the string of regime bits. The use of run-length encoding of the regime automatically allows more fraction bits for the more common values for which magnitudes are closer to 1, and thus provides tapered accuracy in a bit-efficient way. Further details about the posit number format and posit arithmetic can be found in~\cite{john1}. The posit format and IEEE 754-2008 compliant number formats are depicted in Fig. \ref{fig:ieee754}.

In our experiments, we have used IEEE 754 compliant 32-bit (single precision) floating-point numbers and 32-bit posit numbers with $es=2$ that are commonly used. \vspace{-7pt}
\subsection{Soft Errors}
The sharp technology scaling in new generation integrated circuits accentuates the sensitivity of electronic circuits. As a matter of fact, embedded systems are becoming remarkably sensitive to soft errors. These errors result from a voltage transient event induced by alpha particles from packaging material or neutron particles from cosmic rays~\cite{soft_errors}. This event is created due to the collection of charge at a p-n junction after a track of electron-hole pairs is generated. 
In past technologies, this issue was considered in a limited range of applications in which the circuits are operating under aggressive environmental conditions like aerospace applications. Nevertheless, shrinking transistor size and reducing supply voltages in new hardware platforms bring soft errors to ground level mainstream applications \cite{mainstream} \cite{mainstream2}. 

%% file: rw.tex

Since soft errors became a challenging threat to reliability, numerous published work proposes error-resilient memories. Architecture level error resilience techniques such as single error correction double error detection (SECDED) have been proposed and widely used for memory protection \cite{ref5}. The main drawback of SECDED is its area overhead and the supplementary latency leading to performance loss. A fault-tolerant architecture presented in \cite{rsp_ihsen} combines both parity and single redundancy to enhance memories' reliability.
 The weakness of these techniques is their area, power and delay overheads due to the additional memory cells and supporting circuits required for error detection and correction. 

Circuit-level techniques have been proposed to overcome architecture-level overheads. These techniques enhance error resilience at circuit level either by slowing down the response of the circuit to transient events or by increasing its critical charge. Methods such as \cite{ref6} suggest to harden the cell using a pass transistor that is controlled by a refreshing signal.  Hardened memory cells were proposed in \cite{ref7}, \cite{ref16} and \cite{AS8} that add redundant transistors to the 6T-SRAM to increase the cell critical charge. A Schmitt trigger-based technique \cite{ref14} proposes a hardened 13-T memory cell. However, this technique slows down memory due to a Schmitt trigger's hysteresis temporal characteristics. In the context of emerging approximate computing applications, recent works like \cite{smail_prop} proposed a trade-off between reliability and computing precision. To assess the reliability level at an early stage, fault injection can be performed in simulation. All these techniques do not take into account the actual data representation that is stored within the protected memories, especially numerical values.

A number of researchers have approached the reliability issue in numerical algorithms. The vast majority of them treat an algorithm as a black-box and track the behavior of these applications when running with injected soft errors. In \cite{fp_err}, a study on soft error propagation in floating-point programs is presented. In \cite{REF_19}, the behavior of various Krylov methods is analysed. The authors track the variance in iteration count based on the data structure that experiences the bit flip. Authors in \cite{REF_21} analyzed the impact of bit flips in a sparse matrix-vector multiply (SpMV). Exemplifying the concept of black-box analysis of bit flips, \cite{REF_26} presents BIFIT for characterizing applications based on their vulnerability to bit flips.

While these techniques study the reliability of applications based on floating-point formats, none of them study the inherent sensitivity level of floating-point representations. 
This paper proposes a comparative study of the inherent sensitivity to errors in IEEE-754 compliant floating-point and posit representations.

%% file: methodology.tex
We cover analyses for SEU and MBU considering different aspects. Since float and posit have different data formats as shown in Fig. \ref{fig:ieee754}, a single or multiple bit flip event in a 32-bit number results in a new different number for both formats. In our analyses, we consider bit flips in fraction, and exponent for both formats as well as regime bits for posit numbers. For our theoretical analyses, we use numbers $f_1,p_1 \in \R$, where $f_1$ is compliant to IEEE 754-2008 and $p_1$ is a posit number, and $1\le s_1,s_2 \le 32$, $s_1 \in \N$. $s_1$,$s_2$ are the positions of the bit flips in $f_1$ and $p_1$, $s_1 \neq s_2$. For both the number formats, we assume that the SEU and MBU occur at the same position.

\vspace{-7pt}
\subsection{Fraction bits}\label{sec:frac}
\begin{figure}[!t]
\centering
\includegraphics[width=\columnwidth]{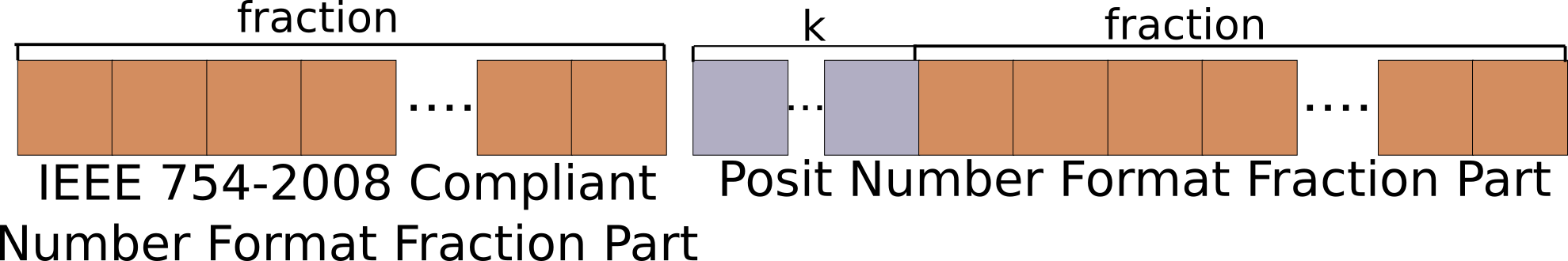}
\caption{Fraction bits in IEEE 754-2008 compliant number and posit compliant number}
\label{fig:frac}
\end{figure}

The total number of fractional bits are $23$ in IEEE 754-2008 and $23+m$ in posit compliant number respectively. $m$-bits are appended in the fraction part in a posit compliant numbers to the left of the fraction bits. Let be $b_{22}b_{21}b_{20},...b_{0}$ and $a_{22+m}, a_{21+m},...a_{0}$, two binary numbers that represent fraction parts of an IEEE 754 compliant number and a posit compliant number respectively. A representative diagram to understand the bit flip phenomena in the fraction part is shown in Fig. \ref{fig:frac}.

The largest error that can occur in IEEE 754-2008 compliant number due to a bit flip in the fraction part is the bit flip of $b_{22}$ ($s_1 = 23$). A flip from $1$ to $0$ or vice-versa would result in subtraction or addition of $0.5$ in the decimal value of the fraction part. On the other hand, in posit to have the similar impact, $s_1$ has to be at $23+m$ bit position. In general, a bit flip in the fraction part of IEEE 754-2008 compliant number is $s_1$ then similar impact in the fraction part of a posit compliant number can be observed if there is a bit flip in the position $s_1+m$. The value of $m$ depends on the configuration of the posit number. For example, a $32$-bit posit number can have k = $-31$ to $30$ regime bits since regime bits are calculated based on run-length of '0' or '1' from the most significant bits after the sign bit (refer equation \ref{eqn:posit}. In practical scenarios the run-length of '0' or '1' is not expected to be very large since a large $k$ results in a very high dynamic range for the numbers. $k=5$ and $es=2$ configuration results in the dynamic range that is similar to the IEEE 754 compliant number. In general, $m = exp\_size - k - es$ where $exp\_size$ is the exponent size in IEEE 754 compliant number and $es$ is the posit exponent size. Since, in the most realistic scenarios $exp\_size > (k+es)$, the bit flip in the position $s_1$ in IEEE 754 compliant number and posit number would result in smaller error in the posit number.     

In case of double bit-flip, the second bit flip position being $s_2$, assuming that the second bit flip occurs in the same locations in an IEEE 754 compliant number and a posit number, the error due to the second bit flip is higher in the IEEE 754 compliant number. The higher error is due the the higher weight associated with the bit position in IEEE 754 compliant number compared to the posit number.  
\vspace{-7pt}

\subsection{Exponent and regime bits}\vspace{-4pt}

A single bit flip in the exponent of IEEE 754 compliant numbers and posit numbers injects a higher error impact in the IEEE 754 compliant number due to the phenomena explained in Section \ref{sec:frac} is applicable to exponent bits as well. Due to more weight associated with the position in the exponent part of the IEEE 754 compliant number compared to the exponent part of the posit number, the error incurred is higher in the IEEE 754 compliant number. A bit flip in the regime part of posit incurs higher error compared to the bit flip in the bits $3$ to $8$ of the float section due to higher weight associated with the posit number. Similarly, second bit flip in the exponent incurs lower error in posit compared to an IEEE 754 compliant number while the second bit flip in regime section of posit number results in higher error. In the subsequent section, we present toolflow to validate our claims.

\subsection{Toolflow}
\begin{figure}
    \centering
     \includegraphics[width=\columnwidth]{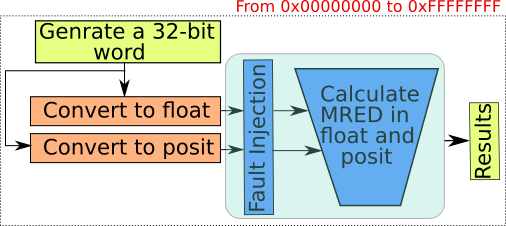}
      \caption{Toolflow of fault injection process in both posit (32,2) configuration and IEEE-754 compliant single-precision floating-point formats (open-source available in \cite{git_fault} }
     \label{fig:flowchart}
 \end{figure}

To assess the impact of errors on both posit and IEEE 754-2008 compliant representations, we proceed to an exhaustive fault injection exploration process. We modified the posit public implementation \cite{softposit} to support our fault injection mechanism. Besides, we built an exhaustive exploration platform shown in Fig. \ref{fig:flowchart}. The idea is to focus on the actual arithmetic representation of the data instead of a coarse grain probabilistic study or a very fine grain circuit simulation.

 \begin{figure*}[!ht]
    \centering
     \subfigure[]{\includegraphics[width=0.85\columnwidth]{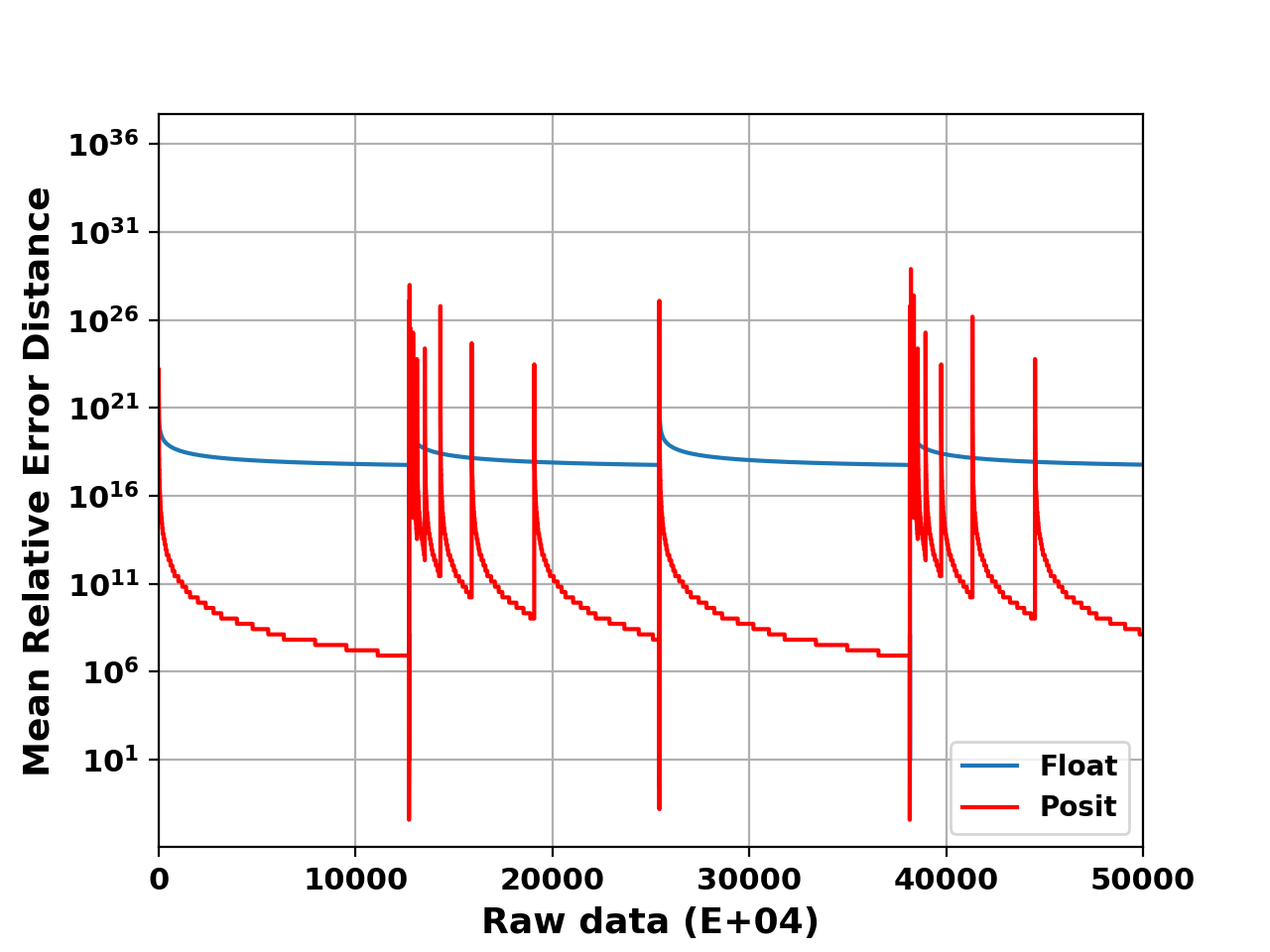}\label{fig:comparaison}}
     \subfigure[]{\includegraphics[width=0.85\columnwidth]{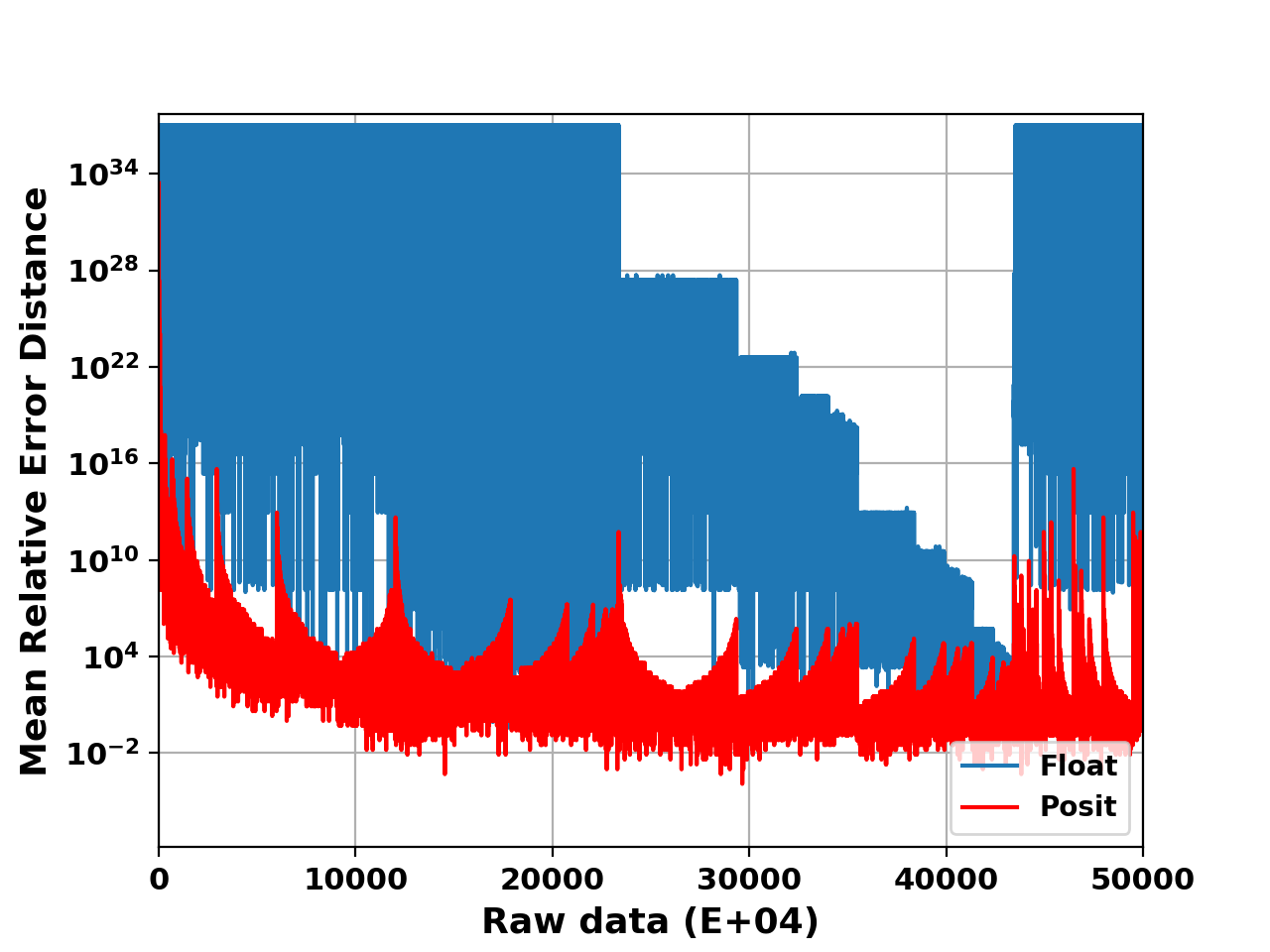}\label{fig:double}}
     \caption{(a)  Mean Relative Error distance comparison between posit and IEEE 754 compliant float under single event upset injection (b) Error distance comparison between posit and IEEE-754 compliant float under double event upset injection}
 \end{figure*}

Fig. \ref{fig:flowchart} explains the followed methodology to assess the inherent reliability of the two tested representations. Since we are considering reliability from a hardware perspective, we are sticking to the actual bit-level data representation.
In this paper, we focus on IEEE 754 single precision floating point and $(N,es) = (32,2)$ (where $N$ is width of the representation and $es$ is exponent size) posit representations for our experiments.
Hence, from a raw 32-bit word, we generate the corresponding floating-point and posit numbers. For a fair comparison, the errors are injected exactly in the same respective bit of the two tested representations. For double bit upsets as well, we choose the same locations for bitflips in both representations.
The inherent reliability of the two representations is assessed by quantifying the mean relative error distance (MRED) of a corrupted value from a golden (non-corrupted) value as shown in Equation \ref{eq_ed}. 
\begin{equation}
\label{eq_ed}
    MRED=\frac{1}{32}* \sum_{i=0}^{31} \frac{|V_i-V_i^*|}{V_i}
\end{equation}
Where $V_i$ and $V_i^*$ are the golden and the corrupted value respectively when a fault is injected in a bit $i$. MRED gives an insight on the mean impact of bit flips that are injected in all words' 32 bits exhaustively.

%% file: results.tex
\begin{figure*}[!t]
\centerline{\includegraphics[width=1.8\columnwidth]{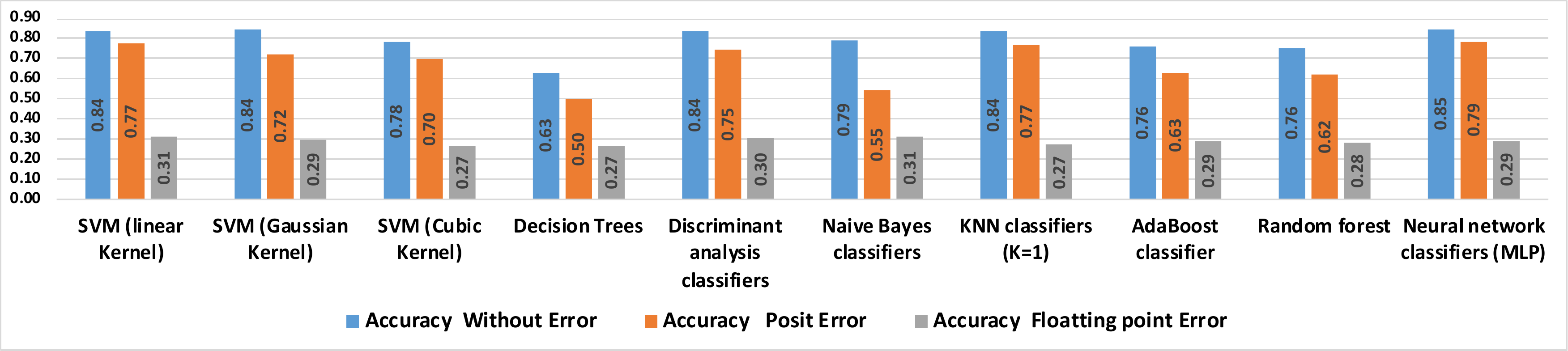}}
\caption{Human action recognition rates using statistical features}
\label{fig:acc_stat}
\end{figure*}

\begin{figure*}[!t]
\centerline{\includegraphics[width=1.9\columnwidth]{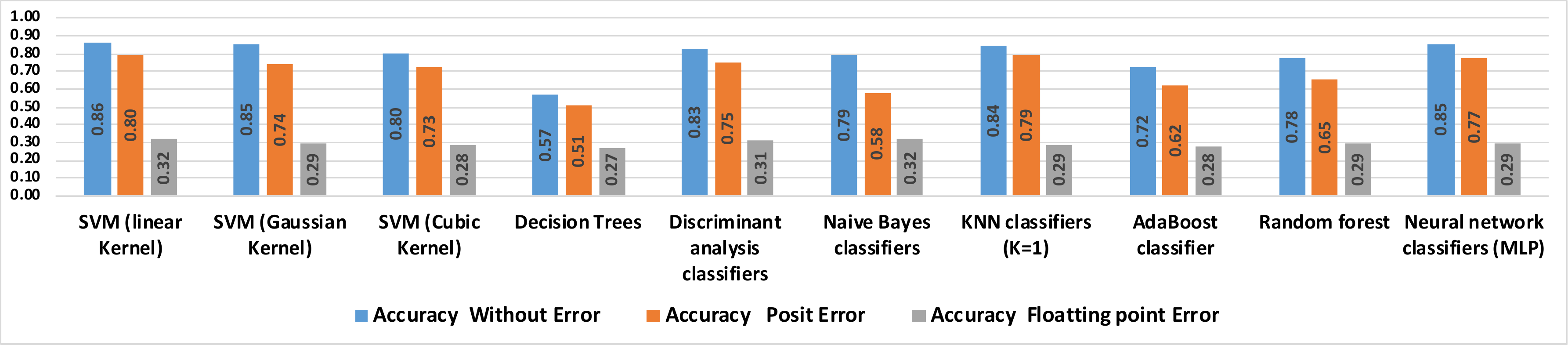}}
\caption{Human action recognition rates using wavelet features.}
\label{fig:acc_wav}
\end{figure*}
The experiments are divided into two categories: 
\begin{itemize}
    \item The first is a comparative application-agnostic exploration of the fault injection impact on reliability of posit and IEE-754 compliant representations. 
    \item The second is a comparative reliability study on a set of ML systems on two different applications tested under fault injection.
\end{itemize}
This section details the experimental setup and discusses the results.
\subsection{Exhaustive comparative reliability exploration}
\label{sec:res_exh}

This set of experiments follow the methodology presented in Section \ref{sec:prop}. The toolflow is implemented in C using the SoftPosit platform \cite{softposit} for posit and a developed bit-wise fault injection platform for IEEE 754 compliant numbers. 
The experiments are run on a 3 GHz Intel Core i7 processor running the OS X 10.9.5 operating system.

\subsubsection{Single Event Upset} \label{sec:res1}
The above-explained experimental setup aims at exploring the impact of bit flips on a given numerical data representation in an exhaustive manner. The results shown in Fig. \ref{fig:comparaison} expose in a logarithmic scale the comparison between posit and IEEE 754 compliant floating-point representations' inherent resiliency to bit flips. The comparison is performed based on the MRED between the golden value (without fault injection) and the corrupted one in both posit and IEEE 754 floats. The results shown in Fig. \ref{fig:comparaison} represent a geometric superposition where the IEEE 754 compliant floating-point graph is in most of the cases above the posit graph. This indicates that the posit representation is globally more error resilient than IEEE 754 compliant representation. In fact, in more than 95\% of the explored cases, a bit flip in an IEEE 754 compliant number deviates from the golden data more than the posit number. Moreover, we registered only 31 cases of not a real (NaR) with posit, which represents 0.7E-6\% of the fault injections. On the other hand, for IEEE compliant floating-point, more than 4\% of the fault injections resulted in not a number (NaN). These cases correspond to non-representable data in the IEEE 754 compliant floating-point graph of Fig. \ref{fig:comparaison}.

\begin{figure*}[!t]
\centerline{\includegraphics[width=1.9\columnwidth]{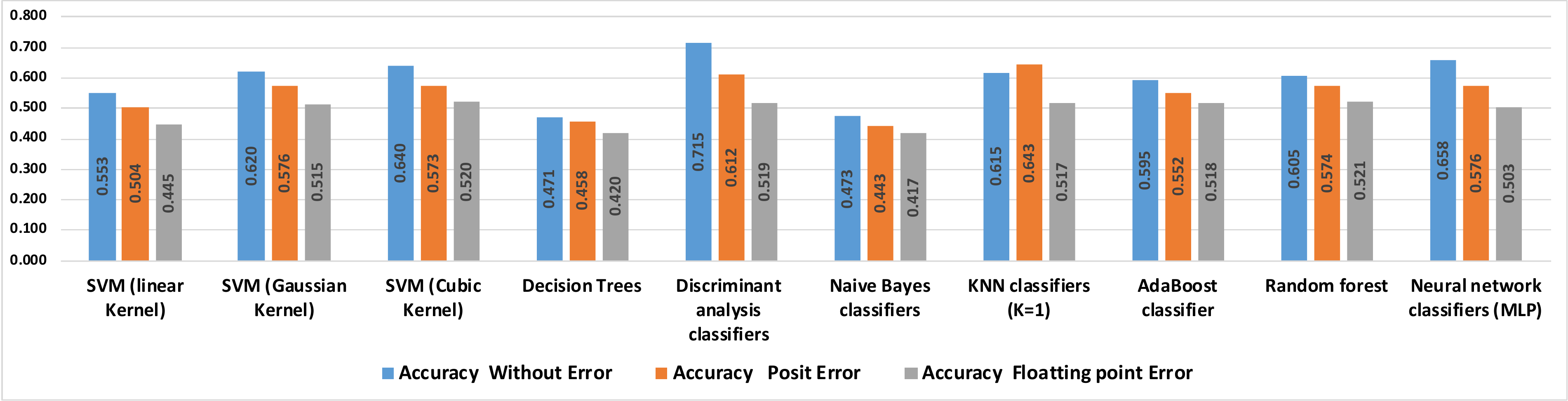}}
\caption{Biometric ECG authentication rates using statistical features.}
\label{fig:ecg_stat}
\end{figure*}

\begin{figure*}[!t]
\centerline{\includegraphics[width=1.9\columnwidth]{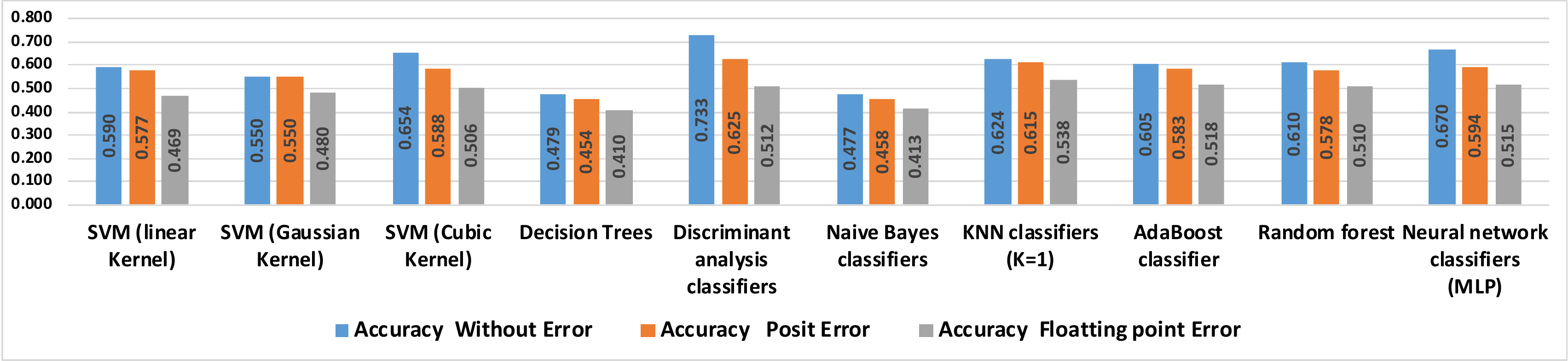}}
\caption{Biometric ECG authentication rates using wavelet features.}
\label{fig:ecg_wav}
\end{figure*}

\subsubsection{Double Event Upset}
Starting from 40nm technology,  more than 35\% of bit upsets are MBUs. Therefore, it is important to consider this phenomenon in reliability assessment processes. In this section, we track the impact of double bit upsets on the data representation for both posit and IEEE 754 compliant floating-point representations. 

Following the same fault injection exploration mechanism as shown in Fig.~\ref{fig:flowchart}, we evaluate the impact of two-bit flips on the MRED between posit and IEEE-compliant floating point. We inject two bit flips at every iteration: the first is injected exhaustively bit-wise, and the second location is randomly selected following a normal distribution among the remaining bits. Fig. \ref{fig:double} shows the MRED caused by two-bit upsets in both representations. Injecting 2 bit flips results globally in higher error magnitudes. However, the results still confirm the higher inherent error resilience of posit shown with the single bit upset experiments. 

The error resilience in posit is due to two main reasons: the variable size of the scale factor (regime bits) and a larger number of bits in the fractional part for the vast majority of cases. The larger number bits in the fractional part is due to variable-sized regime bits. An SEU or MBU in the fractional part results in a lower error compared to an SEU or MBU in the regime bits or exponent bits in posit. Since the IEEE 754-compliant representation has more exponent bits compared to the regime bits and the exponent bits in the posits, the resulting error is higher in IEEE 754 compliant floating-point numbers compared to posits. Better error resilience of posit data-type makes it the right choice for mission-critical next-generation systems. Moreover, the absence of redundant representations such as NaNs is a supplementary factor that enhances posit robustness to errors.
\subsection{Machine-learning applications}
Recent attacks on ML applications are based on deliberate fault-injection techniques \cite{neuro}. In this subsection, we show the results of  fault injection experiments applied in a set of ML applications.
We evaluate two computer-vision systems. The first is a biometric authentication system using Electrocardiogram (ECG) signals based on LATIS ECG Database~\cite{latis}.
The second is a human action recognition (HAR) system using kinematic accelerometer signals trained with Berkley MHAD dataset \cite{mhad}. For the features extraction phase, two types of features were chosen:
\begin{itemize}
    \item Temporal features such as the mean, the standard deviation, the quadratic mean, and the covariance.
    \item Time-frequency type characteristics resulting from the Wavelet transformation.
\end{itemize}

We used the sliding window method to extract the characteristics of each window. These characteristics are subsequently concatenated in a descriptor vector. 

For the classification phase, we evaluate a set of most widely used classifiers in the literature. These ML techniques are: the Support Vector Machines (SVM) with linear, Gaussian and Cubic kernels, Decision Trees, Discriminant analysis classifiers, Naive Bayes classifiers, KNN classifiers (K = 1), AdaBoost classifier, Random forest and Neural network classifiers.
Figures \ref{fig:acc_stat}, \ref{fig:acc_wav}, \ref{fig:ecg_stat} and \ref{fig:ecg_wav} show the recognition rate of the different techniques and settings with and without fault injection. These figures show the impact of single fault injection in the input of the different classifiers for both IEEE floating point and posit data representation. In all these cases with varying features and classifiers, the fault injection impact is significantly lower on the posit implementation which confirms the findings in Section \ref{sec:res_exh}. In fact, while the overall accuracy drop in posit under fault injection varies from $0\%$ and $30\%$, the accuracy drop in IEEE-compliant floating point varies between $10\%$ and $67\%$.

%% file: conclusion.tex
This paper investigates the reliability of two prominent data representations, namely IEEE 754 compliant single precision and (32,2) posit representation. Firstly, we presented a brief theoretical analysis of both the number formats for a single bit flip and double bit flip. An exhaustive fault injection platform is implemented, and the exploration led to a promising conclusion for posit arithmetic that corroborated to our theoretical analysis. To further illustrate this finding, we conduct a benchmark of several ML techniques under fault injection. The experiments demonstrate higher inherent robustness of posit compared to the classical IEEE 754 representation. These findings are useful for safety-critical systems design. They can also be exploited for limiting imprecision in approximate computing designs. Future work will tackle the implementation of a full posit-based processor architecture.